\documentclass[twoside]{article}
\usepackage{amsmath,amssymb,epsfig,float,theorem}
\usepackage[round]{natbib}
\let \Pr \relax
\DeclareMathOperator{\Pr}{\mathbb{P}}
\newcommand{\Prof}[1]{\Pr \left( #1 \right)}
\DeclareMathOperator{\Ex}{\mathbb{E}}
\newcommand{\Exof}[1]{\Ex \left[ #1 \right]}
\newcommand{\PE}{\mathcal{P}_E}
\newcommand{\scf}{\mathsf{s}}
\newcommand{\entr}{\mathsf{e}}
\renewcommand{\div}{\mathsf{d}}
\newcommand{\dd}{\mathrm{d}}
\newcommand{\beq}[1]{\begin{equation}\label{#1}}
\newcommand{\eeq}{\end{equation}}
\title{Reliability, Sufficiency, and the Decomposition of Proper Scores}
\author{Jochen Br\"{o}cker\thanks{email: \texttt{broecker@pks.mpg.de}}\\Max--Planck--Institut f\"{u}r Physik komplexer Systeme\\N\"{o}thnitzer Strasse~34\\01187 Dresden\\Germany}
\begin{document}
\maketitle
\begin{abstract}
Scoring rules are an important tool for evaluating the performance of probabilistic forecasting schemes.
In the binary case, scoring rules (which are strictly proper) allow for a decomposition into terms related to the resolution and to the reliability of the forecast.
This fact is particularly well known for the Brier Score. 
In this paper, this result is extended to forecasts for finite--valued targets.
Both resolution and reliability are shown to have a positive effect on the score.
It is demonstrated that resolution and reliability are directly related to forecast attributes which are desirable on grounds independent of the notion of scores. 
This finding can be considered an epistemological justification of measuring forecast quality by proper scores.
A link is provided to the original work of \citet{degroot82}, extending their concepts of sufficiency and refinement.
The relation to the conjectured sharpness principle of \citet{gneiting05-2} is elucidated.
\end{abstract}
% 
% \keywords{probabilistic forecasts; scoring rules; reliability; resolution}
% 
% \maketitle
% 
\section{Introduction}
\label{sec:introduction}
\citet{brown70} argues that it seems reasonable to value forecasts (be they probabilistic or other) by a scheme related to the extend to which the forecasts ``come true''.
Scoring rules provide examples for such schemes in the case of probabilistic forecasts. After pioneering work by \citet{good52,brier50}, scores were thoroughly investigated in the 1960's and 1970's.
The score was effectively thought of as a reward system, inducing (human) experts to provide their judgments or predictions regarding uncertain events in terms of probabilities~\citep{brown70,savage71}.
In this respect, scoring rules were devices to elicit probabilities from humans.
The importance of using proper scores was recognized already by~\citet{brier50}~\citep[see also][for an entertaining discussion and ``some horrible examples'']{brown70}.
The central argument is that a forecaster's probability assignment should be independent of the particular reward system, which is guaranteed if the reward system constitutes a proper score.
\citet{savage71}~\citep[following][]{definetti70} points out that this universality property allows for an alternative definition of subjective probability, which is a concept of probability independent of the notion of relative observed frequency.

Owing to the enormous increase in computer power over the last decades, it became computationally feasible to numerically produce probabilistic forecasts for dynamical processes, employing models of ever increasing complexity.
Since it is obviously irrelevant whether probabilities are produced by humans or machines, scores provide a tool to evaluate probabilistic numerical forecasting systems, too.
In weather forecasting, scores had already been used to evaluate subjective forecasts (issued by expert meteorologists), for example of rain, long before numerical weather forecasts became available~\citep{brier50,winkler68,epstein69,murphy77}.
Nowadays, scores are widely applied also in the evaluation of numerically generated probabilistic weather forecasts~\citep{gneiting04-2,gneiting04,broecker04,raftery05,wilks06,broecker06-5}.
In contrast to the expert--judgment--forecasts considered in earlier works on scores, weather forecasts are often issued over a long period of time under (more or less) stationary conditions, allowing for archives of forecast--observation pairs to be collected. 
This fact allows to reconsider the interpretation of probabilities as long time observed frequencies.
If we were to forecast the probability of rain on a large number of occasions, we would like rain to occur on a fraction $p$ of those instances where our forecast was (exactly or around) $p$. 
A forecast having this property (up to statistical fluctuations) is called {\em reliable}~\citep{murphy77,toth02,wilks06-2}.
If a large archive of forecast--observation pairs is available, reliability becomes a sensible property to ask for.
As has been widely noted previously though, it is not difficult to produce reliable forecasts if no constraint is put on the information content or resolution of the forecast (the exact meaning of these terms is often left vague, though).
In any event, the grand probability (aka climatological frequency) of the target will always be a reliable forecast, and despite the difficulties with the term ``information content'', many people would presumably agree that this forecast is not very informative.
But how do these virtuous forecast attributes pertain to proper scores?
Do proper scores reward reliable forecasts?
Does a ``better informed'' forecaster really achieve a better score?
In this paper, these questions are answered in the affirmative (using the appropriate formalisation of ``better informed'').
In Section~\ref{sec:decomposition}, after recalling the notion of reliability, it is shown that proper scores allow for a decomposition into terms measuring the resolution and the reliability of the forecast.
In particular, reliability turns out to have a direct positive impact on the score.
In Section~\ref{sec:general-decomposition}, the concept of sufficiency is introduced, generalising similar notions of~\citet{degroot82}.
Sufficiency formalises the idea of ``being more informed'', and is shown to have a direct positive impact on the resolution term of the score. 
The decomposition of Section~\ref{sec:decomposition} is well known for the Brier score~\citep[see for example][]{murphy87,murphy96,lad85}, a widely used score for forecasting problems with only two categories.
The Brier score presumably owes much of its popularity to this decomposition, rendering its interpretation very clear.
\citet{degroot82} have derived a similar decomposition for any proper score in the case of binary targets.
(This result seems not to be widely known in the atmospheric sciences community, and I became aware of it rather belatedly during the preparation of this manuscript.)
The relation to the conjectured sharpness principle of \citet{gneiting05-2} is elucidated.
The appendix contains several more technical points.
Appendix~\ref{apx:reliability} provides an equivalent characterisation of reliability.
In Appendix~\ref{apx:degroot}, the equivalence between sufficiency according to~\citet{degroot82} and as used in this paper is shown.
Finally, the derivation of the decomposition~\eqref{equ:160} is presented in Appendix~\ref{apx:resolution}.
% 
%
% 
%%%%%%%%%%%%%%%%%%%%%%%%%%%%%%%%%%%%%%%%%%%%%%%%%%%%%%%%%%%%%%%%%%%%%%%%%%%%
% 
\section{A general decomposition}
\label{sec:decomposition}
In this section, a general decomposition of proper scores will be derived. To facilitate the discussion, some convenient notation will be introduced first, supplemented with a brief reminder on proper scores.
Let $Y$ denote the quantity to be forecast, commonly referred to as the {\em observation} or {\em target}.\footnote{%
I use italics to indicate that an expression is to be considered a technical term.}
The observation $Y$ is modelled here as a random variable taking values in a set $E$.
For the sake of simplicity, $E$ is assumed to be a finite set of alternatives (e.g.\ ``rain/hail/snow/sunshine''), labelled $1 \ldots K$.
Values of $Y$ (i.e.\ elements of $E$) will be denoted by small lowercase letters like $x$, $y$, or $z$.
A {\em probability assignment} over $E$ is a $K$--dimensional vector $p$ with nonnegative entries so that $\sum_{k \in E} p_k = 1$.
The set of all probability assignments over $E$ is denoted by $\PE$.
Elements of $\PE$ will be denoted by $p, q$, and $r$.
A {\em probabilistic forecasting scheme} is a random variable $\gamma$ with values in $\PE$.
In other words, the realisations of $\gamma$ are probability assignments over $E$.
The reason for assuming $\gamma$ to be random is that forecasting schemes usually process information that will become available before and at forecast time.
For example, if $\gamma$ is a weather forecasting scheme with lead time 48h, it will depend on weather information down to 48h prior to when the observation $Y$ obtains.
The task of designing a forecasting scheme is effectively to model the relationship between this side information and what is to be forecasted~\citep[see][for a related discussion]{murphy87,murphy93,murphy96}.\footnote{%
I do not consider forecasting problems which are explicitly dependent on time, for example to take into account seasonal effects.}
It was already mentioned what reliability means in case that $E$ contains only two elements ($1$ and $0$, say).
In the case of more than two alternatives, this definition of reliability generalises as follows: On the condition that the forecasting scheme is equal to, say, the probability assignment $p$, the observation $Y$ should be distributed according to $p$, or in formulae
\beq{equ:20}
\Prof{Y = k| \gamma = p} = p_k
\eeq
for all $k \in E$.
In particular, a reliable forecasting scheme can be written as a conditional probability. 
As is demonstrated in Appendix~\ref{apx:reliability}, the reverse is also true: every conditional probability of $Y$ is reliable. 
In view of Equation~\eqref{equ:20}, I will fix the notation $\pi^{\gamma}_k := \Prof{Y = k| \gamma}, \, k = 1\ldots K$ for the conditional probability of the observation given the forecasting scheme.
Like every conditional probability, $\pi^{\gamma}$ is a random quantity.
Hence, $\pi^{\gamma}$ is a probabilistic forecasting scheme like $\gamma$ itself.
In terms of $\pi^{\gamma}$ and $\gamma$, the reliability condition~\eqref{equ:20} can be written simply as $ \pi^{\gamma} = \gamma$.
Since $\pi^{\gamma}$ is reliable, it trivially holds that $\pi^{\gamma} = \pi^{\left(\pi^{\gamma}\right)}$.
In any case, $\pi^{\gamma}$ is a function of $\gamma$, independent of whether $\gamma$ is reliable or not.
Let us turn our attention to {\it scoring rules}~\citep[see for example][]{matheson76,gneiting04}. 
A scoring rule is a function $S(p, y)$ which takes a probability assignment over $E$ as its first argument and an element of $E$ as its second argument.
For any two probability assignments $p$ and $q$, the {\em scoring function} is defined as 
\beq{equ:40}
\scf(p, q) = \sum_{k \in E} S(p, k) q_k.
\eeq 
The interpretation of the scoring function is that if $Z$ is a random variable of distribution $q$, then $\scf(p, q)$ is the mathematical expectation of the score of the assignment $p$ in forecasting $Z$.
It is our convention that a small score indicates a good forecast.
A score is called {\em proper} if the {\em divergence}
\beq{equ:50}
\div(p, q) = \scf(p, q) - \scf(q, q)
\eeq 
is nonnegative, and it is called  {\em strictly proper} if $\div(p, q) = 0$ implies $p = q$.
The interpretation of $\div(p, q)$ as a divergence is obviously meaningful only if the scoring rule is strictly proper.
From now on, I assume that scoring rules are strictly proper. 
It is important to note that $\div(p, q)$ is, in general, not a metric, as it is neither symmetric nor does it fulfil the triangle inequality.
The quantity
\beq{equ:60}
\entr(p) = \scf(p, p)
\eeq 
is called the {\em entropy} of $p$.\footnote{%
\citet{gneiting04} refer to $-\entr(p)$ as either the generalised entropy function or the information measure, but since entropy is commonly interpreted as a {\em lack} of information, I define $\entr(p)$ to be the entropy.} 
Table~\ref{tab:scores} gives a couple of frequently used scoring rules along with the corresponding divergences and entropies.
\begin{center}
\texttt{Table~\ref{tab:scores} on top of this or the next page}
\end{center}
% 
%%%%%%%%%%%%%%%%%%%%%%%%%%%%%
% 
\begin{table*}
\begin{minipage}{\linewidth}
\begin{center}
\small
\begin{tabular}{cccc}
Name 
& scoring rule $S(p, y)$ 
& divergence $\div(q, p)$ 
& entropy $\entr(p)$ \rule[-1ex]{0mm}{3.5ex} \\\hline

Brier\footnote{%
For binary cases (i.e.\ $E = \{ 0, 1 \}$).}
& $| y - p|^2$
& $|p - q|^2$
& $p (1 - p)$ \rule[-2ex]{0mm}{5ex}\\
Ignorance\footnote{%
Propriety follows from Jensen's inequality.}
& $-\log p_y$
& $\sum -\log \left( \frac{p_k}{q_k} \right) q_k $
& $\sum -\log (p_k) p_k $  \rule[-2ex]{0mm}{5ex}\\
CRPS\footnote{%
Continuous Ranked Probability Score -- Here $F$ and $G$ are the cumulative distribution functions corresponding to $p$ and $q$, respectively.}
& $ \int (F(z) - H(y - z))^2 \dd z$
& $ \int (F(z) - G(z))^2 \dd z$
& $ \int F(z)(1 - F(z)) \dd z$  \rule[-2ex]{0mm}{5ex}\\
PSS\footnote{%
Pseudo-spherical Scores -- Here $\alpha > 1$, while
$\| p \|_{\alpha} = \left[ \sum p_k^{\alpha} \right]^{1/\alpha}$.
Propriety follows from H\"{o}lder's Inequality.}
& $-\frac{p_y^{\alpha-1}}{\| p \|^{\alpha-1}_{\alpha}}$
& $\| q \|_{\alpha} - \frac{\langle q, p(z)^{\alpha-1} \rangle}{\| p \|^{\alpha-1}_{\alpha}}$
& $ - \| p \|_{\alpha}$  \rule[-2ex]{0mm}{5ex}\\
PLS\footnote{%
Proper Linear Score, also referred to as the quadratic score. For binary cases (i.e.\ $E = \{ 0, 1 \}$), this score is equivalent to the Brier score}
& $\sum p_k^2 - 2 p_y$
& $\sum (p_k - q_k)^2 $
& $- \sum p_k^2 $  \rule[-2ex]{0mm}{5ex}\\
\end{tabular}
\end{center}
% Footnotes appear here
\end{minipage}
\caption{\label{tab:scores}Scoring rule, divergence, and entropy for several common scores. All sums extend over $E$. See \citet{epstein69,murphy70} for a discussion of the Ranked Probability Score. \citet{matheson76,gneiting04} discuss scoring rules for continuous variables.}
\end{table*}
% 
% 
%%%%%%%%%%%%%%%%%%%%%%%%%%%%
% 

%
For strictly proper scores, 
\beq{equ:70}
\entr(p) = \inf_q \scf(q, p).
\eeq 
Since $\scf(q, p)$ is linear in $p$, Equation~\eqref{equ:70} demonstrates that for strictly proper scores, the entropy $\entr(p)$ is an infimum over linear functions and hence concave~\citep{rockafellar70}.
For the particular cases listed in Table~\ref{tab:scores}, it should be fairly obvious that the entropy is a measure for the uncertainty inherent in a probability assignment $p$.
For the Brier score and the Ignorance, the entropy is indeed a very common measure of inherent randomness of a distribution.
Furthermore, suppose $p$ and $q$ are two probability assignments featuring the same entropy, then intuitively, any mixture of $p$ and $q$ should have a larger inherent uncertainty than any of the individual probability assignments, an intuition which the entropy supports, due to the concavity of $\entr(p)$.
Our aim now is to derive a decomposition of the {\em expected score} $\Exof{S(\gamma, Y)}$ of the forecasting scheme $\gamma$.
Since $\gamma$ is random, the expectation affects both $\gamma$ and $Y$.
An elementary property of the mathematical expectation gives
\beq{equ:80}
\Exof{S(\gamma, Y)} = \Exof{\Exof{S(\gamma, Y) | \gamma}}.
\eeq
To calculate the conditional expectation $\Exof{S(\gamma, Y) | \gamma}$, the probability of $Y$ given $\gamma$ is needed, but this is just $\pi^{\gamma}$, whence
\beq{equ:90}
\Exof{S(\gamma, Y) | \gamma} = \scf(\gamma, \pi^{\gamma}).
\eeq
Substituting with Equation~\eqref{equ:90} in~\eqref{equ:80} results in 
\beq{equ:100}
\Exof{S(\gamma, Y)} = \Ex \scf(\gamma, \pi^{\gamma}).
\eeq
From Equations~\eqref{equ:50} and~\eqref{equ:60} we get
\beq{equ:110}
\scf(\gamma, \pi^{\gamma}) = \entr(\pi^{\gamma}) + \div(\gamma, \pi^{\gamma}).
\eeq
Taking the expectation on both sides of Equation~\eqref{equ:110} and substituting for the right hand side in~\eqref{equ:100}, we obtain
\beq{equ:120}
\Exof{ S(\gamma, Y) } = \Ex \entr(\pi^{\gamma}) + \Ex \div(\gamma, \pi^{\gamma}).
\eeq
The first term in Equation~\eqref{equ:120}, the average entropy of $\pi^{\gamma}$, can be decomposed further. 
Consider the (nonrandom) assignment obtained by taking the average of $\pi^{\gamma}$, 
\beq{equ:130}
\bar{\pi} := \Ex \pi^{\gamma}
\eeq
It is easily seen that $\bar{\pi}$ is just the unconditional probability of $Y$, which in meteorology is often referred to as the {\em climatology} of $Y$.
Since $\scf(\bar{\pi}, \pi^{\gamma})$ is linear in $\pi^{\gamma}$ and $\bar{\pi}$ is not random, it follows immediately from Equation~\eqref{equ:130} that 
\beq{equ:140}
\Ex \scf(\bar{\pi}, \pi^{\gamma}) = \scf(\bar{\pi}, \bar{\pi}) = \entr(\bar{\pi}).
\eeq
Adding and subtracting $\Ex \scf(\bar{\pi}, \pi^{\gamma})$ on the right hand side of Equation~\eqref{equ:120} and using Equation~\eqref{equ:140} we arrive at
\beq{equ:150}
\Ex \scf(\gamma, y) = \entr(\bar{\pi}) - \Ex \div(\bar{\pi}, \pi^{\gamma}) + \Ex \div(\gamma, \pi^{\gamma}).
\eeq
Equation~\eqref{equ:150} constitutes the desired decomposition of the expected score of the probabilistic forecasting scheme $\gamma$.
This decomposition is, as I will argue, completely analogous to and a generalisation of the well known decomposition of the Brier score.
The three terms in Equation~\eqref{equ:150} will be (from left to right) referred to as  the uncertainty of $Y$, the resolution term\footnote{Also called sharpness term}, and the reliability term.
As a starting point for the discussion of the decomposition~\eqref{equ:150}, the reader might want to convince himself (with the help of Table~\ref{tab:scores}) that for the Brier score, Equation~\eqref{equ:150} indeed yields the known decomposition.
Firstly, the uncertainty of $Y$ is the entropy of the climatology and hence can be interpreted as the expected score of the climatology as a forecast, quantifying the ability of the climatology to forecast random draws from itself.
The resolution term $\Ex \div(\bar{\pi}, \pi^{\gamma})$ contributes negatively to the score.
Note that due to the strict propriety of the score, the resolution is always positive definite.
Since the resolution term describes the average deviation of $\pi^{\gamma}$ from its average $\bar{\pi}$ (see Equation~\ref{equ:130}), it can be interpreted as a form of variance of $\pi^{\gamma}$.
The larger the variance, the better the score. 
This term reduces to the standard variance of $\pi^{\gamma}$ in case of the Brier score.
Finally, the reliability term (which is again positive definite) describes the average deviation of $\gamma$ from $\pi^{\gamma}$.
Recalling that $\gamma = \pi^{\gamma}$ indicates a reliable forecast, the interpretation of the reliability term as the average violation of reliability becomes obvious.
\section{A decomposition of the resolution term}
\label{sec:general-decomposition}
The decomposition~\eqref{equ:150} demonstrates how the score changes if the forecast scheme $\gamma$ changes, but so that $\pi^{\gamma}$ remains constant. 
In this case, any deviation of $\gamma$ from $\pi^{\gamma}$ has adverse effects on the score.
But in general, changing $\gamma$ means that $\pi^{\gamma}$ changes, too.
Thus, changes $\gamma$ usually entail changes in both the reliability and the resolution term of the decomposition~\eqref{equ:150}.
The changes in the resolution term are investigated in this section.
The intuitive interpretation of the resolution term is that it somehow measures the average information content of the forecast scheme.
In this section, I will discuss the concept of forecast sufficiency, introduced by~\citet{degroot82}. 
This concept formalises the notion of being ``more or less informed'' and allows for the partial ordering of forecasting schemes.
As will be seen in this section, $\gamma_1$ will have at least the same resolution as $\gamma_2$ if $\gamma_1$ if sufficient for $\gamma_2$. 
Thus, the expected score reproduces the same ordering as sufficiency. 
This result establishes a connection between a quantitative notion of information as provided by the score, and a qualitative notion of information contents as provided by sufficiency.
This is analogous to the relation between the reliability term of the decomposition~\eqref{equ:150} and the qualitative reliability condition~\eqref{equ:20}.
I call a forecasting scheme $\gamma_1$ {\em sufficient} for a forecasting scheme $\gamma_2$ if  
\beq{equ:155}
\pi^2 = \Exof{\pi^1 | \gamma^2},
\eeq
where the abbreviations $\pi^1 := \pi^{\gamma_1} = \Prof{Y | \gamma_1}$ and analogously for $\pi^2$ were used.\footnote{%
If both~$\gamma_1$ and~$\gamma_2$ are reliable, then condition~\eqref{equ:155} modifies to $\gamma^2 = \Exof{\gamma^1 | \gamma^2}$. In this situation, $\gamma^1$ is said to be {\em at least as refined as} $\gamma^2$.}
In Appendix~\ref{apx:degroot}, it is shown that the present notion of sufficiency is equivalent to the corresponding definition of~\citet{degroot82}.
Before continuing with score decompositions, let me try to elucidate the rather technical condition~\eqref{equ:155} with a somewhat informal interpretation.
Suppose the forecaster who is running forecasting scheme $\gamma_1$, albeit having no access to the current value of $\gamma_2$, collected a large archive of past values of $\gamma_2$ and hence is able to fit a good approximation to $\Prof{\gamma_2 | \gamma_1}$.
With this information, he tries to mimic forecasting scheme $\gamma_2$ as follows.
The forecaster's mimicry version of $\gamma_2$ (which we denote by $\gamma^*_2$) is just a random draw of $\Prof{\gamma_2 | \gamma_1}$ (conditioned on his own forecast $\gamma_1$).
Since the expected score of any forecast scheme depends only on the compound distribution of the forecast scheme and $Y$, the mimicry forecast $\gamma^*_2$ will achieve the same expected score as the real $\gamma_2$ if the compound distribution of $(\gamma_2, Y)$ and  $(\gamma^*_2, Y)$ are the same. 
It is straight forward to work out that the latter condition is equivalent to~\eqref{equ:155}. 
In brief, if $\gamma_1$ is sufficient for $\gamma_2$, then by appropriate randomisation of $\gamma_1$, a forecast $\gamma^*_2$ is obtained which has the same statistical properties as $\gamma_2$.
Note also that in particular $\gamma_1$ is sufficient for $\gamma_2$ if $\gamma_2$ can be written as a function of $\gamma_1$.
In Appendix~\ref{apx:resolution}, it is shown that if $\gamma_1$ is sufficient for $\gamma_2$, it holds that
\beq{equ:160}
\Ex \div(\bar{\pi}, \pi^2) = \Ex \div(\bar{\pi}, \pi^1)
- \Ex \div(\pi^2, \pi^1).
\eeq
Keeping in mind that $\Ex \div(\bar{\pi}, \pi^1)$ and $\Ex \div(\bar{\pi}, \pi^2)$ are the resolution terms of $\gamma_1$ and $\gamma_2$, respectively, and that $\div(\ldots)$ is never negative, Equation~\eqref{equ:160} demonstrates that the resolution of $\gamma_2$ will be at most that of $\gamma_1$.
More generally, Equations~\eqref{equ:150} and~\eqref{equ:160} together allow for the following conclusions as to the approach of scoring forecasting schemes using strictly proper scores:
\begin{itemize}
\item The forecasting scheme $\pi^{\gamma}$ achieves the best possible average score among all forecasts for which $\gamma$ is sufficient.
If the score is strictly proper, $\pi^{\gamma}$ is uniquely defined through this optimum property, in the sense that any forecast for which $\gamma$ is sufficient is either equal to $\pi^{\gamma}$ or it will have a worse average score. 
This can be considered an answer to the conjectured sharpness principle of \citet{gneiting05-2}, reinterpreted in our framework.
\item Per se, it is impossible to say how the score will rank unreliable forecast schemes, even if one is sufficient for the other.
The lack of reliability of one forecast scheme might be outbalanced by the lack of resolution of the other.
\item It is also not clear how the score will rank forecast schemes (reliable or unreliable) as long as none of the two forecast schemes is sufficient for the other.
It seems plausible that the actual ranking of such forecasts will depend on the particular scoring rule employed.
\end{itemize}
% 
% 
%%%%%%%%%%%%%%%%%%%%%%%%%%%%%%%%%%%%%%%%%%%%%%%%%%%%%%%%%%%%%%%%%%%%%%%%%%%%
% 
\section{Conclusion}
\label{sec:conclusion}
The score of a probabilistic forecast was shown to decompose into terms related to the uncertainty in the observation, the resolution of the forecast, and its reliability, generalising corresponding results for the Brier score.
The only property required of the score is that it be strictly proper.
By using a widely accepted characterisation of reliability, and furthermore by generalising the concepts of sufficiency and refinement due to \citet{degroot82}, it was argued that both the resolution and the reliability term in the decomposition quantify forecast attributes for which the case can been made independently (i.e.\ not referring to scoring rules).
These results provide an epistemological justification of measuring forecast quality by proper scores.
Furthermore, the relation to the conjectured sharpness principle of \citet{gneiting05-2} was mentioned.
%
% 
%%%%%%%%%%%%%%%%%%%%%%%%%%%%%%%%%%%%%%%%%%%%%%%%%%%%%%%%%%%%%%%%%%%%%%%%%%%%
% 
\section*{Acknowledgements}
The author gratefully acknowledges fruitful discussions with Kevin Judd, University of Western Australia, as well as the members of the Time Series Analysis group and the Max--Planck--Institute for the Physics of Complex Systems, in particular Gianluigi Del Magno and Holger Kantz.
% 
% 
%%%%%%%%%%%%%%%%%%%%%%%%%%%%%%%%%%%%%%%%%%%%%%%%%%%%%%%%%%%%%%%%%%%%%%%%%%%%
% 
% 
\appendix
\section*{Appendix}
% 
% 
%%%%%%%%%%%%%%%%%%%%%%%%%%%%%%%%%%%%%%%%%%%%%%%%% 
% 
\section{An alternative definition of reliability}
\label{apx:reliability}
In this section, it will be shown that any conditional probability is reliable.
The reader is assumed to be familiar with the basic notions of probability theory~\citep[see e.g.][chapter~4]{BRE}.
Let $\gamma$ be a probabilistic forecasting scheme which can be written as a conditional probability, that is 
\beq{equ:200}
\Prof{Y = k | \mathcal{F}} = \gamma_k
\eeq   
for all $k \in E$ and some sigma algebra $\mathcal{F}$.
On both sides of Equation~\eqref{equ:200}, we take the mathematical expectation conditioned on $\gamma$.
The right hand side gives back $\gamma_k$.
To compute the left hand side, note that because of Equation~\eqref{equ:200}, $\gamma$ is $\mathcal{F}$--measurable.
Hence
\beq{equ:210}
	\begin{split}
	\Exof{\Prof{Y = k | \mathcal{F}} | \gamma}
	& = \Exof{\Exof{\delta_{Y, k} | \mathcal{F}} | \gamma}\\
	& = \Exof{\delta_{Y, k} | \gamma}\\
	& = \Prof{Y = k |\gamma}.
	\end{split}
\eeq   
This demonstrates that $ \Prof{Y = k |\gamma} = \gamma_k$, which is the condition for reliability.
% 
%%%%%%%%%%%%%%%%%%%%%%%%%%%%%%%%%%%%%%%%%%%%%%%%%%%%%%%%%%%%%%%%%%%%%%%%%%%%
% 
\section{Sufficiency and refinement of DeGroot and Fienberg}
\label{apx:degroot}
Let $\gamma_1, \gamma_2$ and $\pi^1, \pi^2$ as in Section~\ref{sec:general-decomposition}.
With these definitions, $\gamma_1$ is sufficient for $\gamma_2$ if $\pi^{2} = \Exof{\pi^1 | \gamma_2}$.
It will now be shown that this is equivalent to the sufficiency condition given by~\citet{degroot82}, Equation~(4.3).
To state the latter condition, I assume that the conditional probability of $\gamma_1$ given $Y$ and the conditional probability of $\gamma_2$ given $Y$, respectively, have densities $g_1(p| Y)$ and $g_2(p| Y)$, respectively.
Furthermore, the conditional probability of $\gamma_2$ given $\gamma_1$ is assumed to have a density $h(\gamma_2| \gamma_1)$.
With these conventions, $\gamma_1$ is sufficient for $\gamma_2$ in the sense of of~\citet{degroot82}, if
\beq{equ:degroot}
g_2(\gamma_2| Y) = \int_{\mathcal{P}_E}\! h(\gamma_2| \gamma_1) \, g_1(\gamma_1| Y) \, \dd \gamma_1.
\eeq
Multiplying both sides by $\bar{\pi}$ and dividing by the density of $\gamma_2$ we obtain 
\beq{equ:180}
\pi^2(\gamma_2) = \int_{\mathcal{P}_E}\! \pi^1(\gamma_1) \, f(\gamma_1| \gamma_2) \, \dd \gamma_1,
\eeq
with $f(\gamma_1| \gamma_2)$ being the conditional probability of $\gamma_1$ given $\gamma_2$.
Here we need to write explicitely that $\pi^1$ and $\pi^2$ depend on $\gamma_1$ and $\gamma_2$, respectively.
But the right hand side of Equation~\eqref{equ:180} is just $\Exof{\pi^1 | \gamma_2}$.
%
% 
%%%%%%%%%%%%%%%%%%%%%%%%%%%%%%%%%%%%%%%%%%%%%%%%%%%%%%%%%%%%%%%%%%%%%%%%%%%%
% 
\section{Derivation of Equation~\ref{equ:160}}
\label{apx:resolution}
Still, $\gamma_1, \gamma_2$ and $\pi^1, \pi^2$ are as in Section~\ref{sec:general-decomposition}.
By just applying definitions, we get
\beq{equ:170}
	\begin{split}
	\div(\bar{\pi}, \pi^2)
	 & = \scf(\bar{\pi}, \pi^2) - \scf(\pi^2, \pi^2)\\
	 & = \scf(\bar{\pi}, \pi^2) - \scf(\pi^1, \pi^1) \\
	 & \qquad - (\scf(\pi^2, \pi^2) - \scf(\pi^1, \pi^1)).
	\end{split}
\eeq
The mathematical expectation of the first term can be written as
\beq{equ:175}
	\begin{split}
	\Ex \scf(\bar{\pi}, \pi^2)
	 & = \Exof{\Exof{S(\bar{\pi}, Y) | \gamma_2}} \\
	 & = \Exof{\Exof{S(\bar{\pi}, Y) | \gamma_1}} \\
	 & = \Ex \scf(\bar{\pi}, \pi^1),
	\end{split}
\eeq
using elementary properties of the conditional expectation and the fact that $\bar{\pi}$ is not random.
Next, the mathematical expectation of the third term is considered:
\beq{equ:177}
	\begin{split}
	\Ex \scf(\pi^2, \pi^2)
	 & = \Exof{\scf(\pi^2, \Exof{\pi^1 | \gamma_2})}\\
	 & = \Exof{ \Exof{\scf(\pi^2, \pi^1)| \gamma_2}}\\
	 & = \Ex \scf(\pi^2, \pi^1).
	\end{split}
\eeq
The first equality is due to sufficiency; the second is valid because $\pi^2$ is a function of $\gamma_2$, so it can be taken under any expectation conditioned on $\gamma_2$; and the third equality uses elementary properties of the conditional expectation.
Taking the expectation over Equation~\eqref{equ:170} and using Equations~(\ref{equ:175}, \ref{equ:177}), we obtain Equation~\eqref{equ:160}. 
% 
% \bibliographystyle{plainnat}
% \bibliography{/home/broecker/TeX/Literatur}
% 
% 
% 

% 
% 
% 
\end{document}